\documentclass[twocolumn]{25onr_art}
\usepackage{ulem}
\usepackage{harvard}
\usepackage{graphicx}

\pdfoutput = 1

\def\be{\begin{eqnarray}}
\def\ee{\end{eqnarray}}
\def\benl{\begin{eqnarray*}}
\def\eenl{\end{eqnarray*}}

\newcommand{\nwc}{\newcommand}
\nwc{\bm}{\boldmath}
\nwc{\m}{\mbox}
\nwc{\ubm}{\unboldmath}
\nwc{\bmU}{\m{\bm$U$\ubm}}
\nwc{\bmX}{\m{\bm$X$\ubm}}
\nwc{\bmu}{\m{\bm$u$\ubm}}
\nwc{\bmx}{\m{\bm$x$\ubm}}
\nwc{\bmz}{\m{\bm$z$\ubm}}
\nwc{\bmv}{\m{\bm$v$\ubm}}
\nwc{\bmw}{\m{\bm$w$\ubm}}
\nwc{\bmW}{\m{\bm$W$\ubm}}
\nwc{\bmn}{\m{\bm$n$\ubm}}
\nwc{\bmG}{\m{\bm$G$\ubm}}
\nwc{\bmF}{\m{\bm$F$\ubm}}
\nwc{\bmI}{\m{\bm$I$\ubm}}
\nwc{\bmN}{\m{\bm$N$\ubm}}
\nwc{\bmP}{\m{\bm$P$\ubm}}
\nwc{\bmcalP}{\m{\bm $\cal P$\ubm}}
\nwc{\bmV}{\m{\bm$V$\ubm}}
\nwc{\bmS}{\m{\bm$S$\ubm}}

\begin{document}

\title{The numerical simulation of ship waves using cartesian grid methods with adaptive mesh refinement}

\author{Douglas G.\ Dommermuth$^1$, Mark Sussman$^2$, Robert F. Beck$^3$, Thomas T. O'Shea$^1$, Donald C. Wyatt$^1$,
Kevin Olson$^4$, and Peter MacNeice$^5$}

\affiliation{\small $^1$Naval Hydrodynamics Division, Science Applications International Corporation,
\\ 10260 Campus Point Drive, MS 34, San Diego, CA  92121 \\ $^2$Department of Mathematics,
Florida State University, Tallahassee, FL 32306 \\ $^3$Department of Naval Architecture \& Marine Engineering, University of
Michigan, \\ 209 NA\&ME Building, 2600 Draper Road, Ann Arbor, Michigan 48109 \\ $^4$University of Maryland at Baltimore County,
Code 931, NASA/GSFC, Greenbelt, MD 20771, \\ $^5$Drexel University, Code 931, NASA/GSFC, Greenbelt, MD 20771}

\maketitle

\begin{abstract}
Cartesian-grid methods with Adaptive Mesh Refinement (AMR) are ideally suited for simulating the breaking of waves, the formation
of spray, and the entrainment of air around ships. As a result of the cartesian-grid formulation, minimal input is required to
describe the ships geometry.  A surface panelization of the ship hull is used as input to automatically generate a
three-dimensional model. No three-dimensional gridding is required.   The AMR portion of the numerical algorithm automatically
clusters grid points near the ship in regions where wave breaking, spray formation, and air entrainment occur.   Away from the
ship, where the flow is less turbulent, the mesh is coarser. The numerical computations are implemented using parallel algorithms.
Together, the ease of input and usage, the ability to resolve complex free-surface phenomena, and the speed of the numerical
algorithms provide a robust capability for simulating the free-surface disturbances near a ship. Here, numerical predictions, with
and without AMR, are compared to experimental measurements of ships moving with constant forward speed, including a vertical strut,
the DDG 5415, and a wedge-like geometry.
\end{abstract}

\section{Introduction}

Two different cartesian-grid methods have been developed to simulate ship waves.  One technique (CLSVOF) combines Level-Set (LS)
techniques with Volume-of-Fluid (VOF) methods to model the free-surface interface.  The second technique uses a pure VOF
formulation.  The CLSVOF formulation uses Adaptive Mesh Refinement (AMR) to resolve small-scale features in the flow.   The VOF
formulation uses domain decomposition without AMR.  Both methods that are described in this paper use the same panelized geometry
that is required by potential-flow methods to automatically construct a signed-distance-function representation of the hull (see
\citeasnoun{sussman01}). The hull representation is then immersed inside a cartesian grid that is used to track the free-surface
interface.  No additional gridding beyond what is already used by potential-flow methods is required. The CLSVOF formulation is
used to investigate the flow around the DDG 5415, and the pure VOF formulation is used to model the flow around a vertical strut
and a wedge-like geometry.  In all cases, comparisons are made to experiments.

The Numerical Flow Analysis (NFA) code is meant to provide a turnkey capability to model breaking waves around a ship, including
both plunging and spilling breaking waves, the formation of spray, and the entrainment of air.   Cartesian-grid methods are used to
model the ship hull and the free surface. Following \citeasnoun{goldstein93} and \citeasnoun{sussman01}, a body-force method is
used to enforce a no-slip boundary condition on the hull.  Based on \citeasnoun{colella99}, the ability to impose free-slip
boundary conditions is also provided.   A surface representation of the ship hull is used as input to construct a three-dimensional
representation of the ship hull on a cartesian grid.  The interface capturing of the free surface uses a second-order accurate, VOF
technique.   At each time step, the position of the free surface is reconstructed using piece-wise planar surfaces as outlined in
\citeasnoun{rider94}. Based on \citeasnoun{iafrati01}, the in-flow and out-flow boundary conditions use a body-force technique to
enforce a uniform stream with no free-surface disturbance ahead of and behind the ship. A second-order, variable-coefficient
Poisson equation is used project the velocity onto a solenoidal field thereby ensuring mass conservation. A preconditioned
conjugate-gradient method is used to solve the Poisson equation. The convective terms in the momentum equations are accounted for
using a slope-limited, third-order QUICK scheme as discussed in \citeasnoun{leonard97}.   Based on the PARAMESH suite of codes
\cite{macneice00}, domain decomposition is used to solve the field equations.  PARAMESH controls data communication between blocks
of grid points, and also between computer processors. PARAMESH is written in Fortran 90.  PARAMESH provides AMR capability, but
here we only illustrate the NFA code using uniform grid spacing without adaptive meshing.  (An AMR capability for the NFA code is
in progress.) On the Cray T3E, message passing is accomplished using either the Cray SHMEM library or MPI. The CPU requirements are
linearly proportional to the number of grid points and inversely proportional to the number of processors.  For the NFA code,
comparisons are made to measurements of flow around a vertical strut \cite{zhang96}and a wedge-like geometry \cite{karion03}.

Developed concurrently with the NFA code, another code based on the Coupled Level set and Volume-of-Fluid (CLSVOF) method has been
developed for modelling free-surface flows in general geometries.   The CLSVOF code uses adaptive mesh refinement to compute
multi-scale phenomena.  Like the NFA code, the CLSVOF code uses cartesian grid techniques to model complex geometries.  Also, like
NFA, CLSVOF uses a two-phase formulation of the air-water interface.  Unlike the NFA code, which is based on PARAMESH, the CLSVOF
code is based on BOXLIB, which is developed by the CCSE group at Lawrence Berkeley National Laboratories. The strategy of BOXLIB is
that high-level adaptive gridding and parallel functions are performed using C++ while numerical discretizations of the
Navier-Stokes equations are performed using a FORTRAN code.  The BOXLIB libraries take care of all the dynamic gridding functions,
whereas the user only has to supply FORTRAN routines that operate on fixed, uniform, rectangular grids. Please refer to the work of
\citeasnoun{rendleman00} for more information regarding BOXLIB.  For computation of incompressible flow on an adaptive grid, it is
not enough to insure that fluxes are matched at coarse/fine grid boundaries.  We must also compute a "composite'' projection step
at each time step.  A "composite'' projection step insures that the pressure, velocity, and divergence-free condition, are
satisfied across coarse-fine grid boundaries. For details of our adaptive implementation, we refer the reader to
\citeasnoun{sussman03b} and the references therein.  CLSVOF predictions are compared to measurements of the flow around the DDG
5415 (see http://www50.dt.navy.mil/5415/).

CLSVOF and VOF formulations pose unique challenges associated with data processing of the free-surface interface. These challenges
are not unlike those facing experimentalists in the laboratory and in the field. In particular, techniques are required to analyze
unsteady effects, including the formation of bubbles and spray.  Here, we propose various statistical approaches.  For example, two
approaches are proposed for analyzing the free-surface elevation as predicted by VOF formulations.  The first technique
reconstructs the free surface using the average of the volume fractions over time.  The second technique takes the mean and
variance over time of the "zero-crossings'' throughout a column of fluid.   The first technique is useful for predicting the mean
surface elevation. In addition, it provides a prediction of air entrainment beneath the surface and droplet formation above the
surface.   The second technique forms the basis for investigating the variance in the free-surface elevation.  Generally speaking,
high variance indicates regions where either bubbles are entrained or droplets are shed.  These approaches and their nuances are
discussed in greater detail in the results section.

\section{\label{sec:formulation}Formulation}

Consider turbulent flow at the interface between air and water.  Let $u_i$ denote the three-dimensional velocity field as a
function of space ($x_i$) and time ($t$).  For an incompressible flow, the conservation of mass gives
\begin{eqnarray}
\label{mass}
\frac{\partial u_i}{\partial x_i} = 0 \;\; .
\end{eqnarray}
\noindent  $u_i$ and $x_i$ are normalized by $U_o$ and $L_o$, which denote the free-stream velocity and the length of the body,
respectively.

Following a procedure that is similar to \citeasnoun{rider94}, we let $\phi$ denote the fraction of fluid that is inside a cell. By
definition, $\phi=0$ for a cell that is totally filled with air, and $\phi=1$ for a cell that is totally filled with water.

The convection of $\phi$ is expressed as follows:
\begin{eqnarray}
\label{vof}
\frac{d \phi}{d t} = \frac{\partial Q}{\partial x_j} \;\; ,
\end{eqnarray}
\noindent where $d/dt=\partial/\partial t + u_i \partial/\partial x_i$ is a substantial derivative. $Q$ is a sub-grid-scale flux
which can model the entrainment of gas into the liquid.  Details are provided in \citeasnoun{dommermuth98}.

\newpage
Let $\rho_\ell$ and $\mu_\ell$ respectively denote the density and dynamic viscosity of water. Similarly, $\rho_g$ and $\mu_g$ are
the corresponding properties of air.  The flows in the water and the air are governed by the Navier-Stokes equations:
\begin{eqnarray}
\label{navi}
\frac{d u_i}{d t} & = & F_i -\frac{1}{\rho} \frac{\partial
P}{\partial x_i} +\frac{1}{\rho R_e} \frac{\partial}{\partial x_j} \left( 2 \mu S_{ij} \right)
\nonumber \\ & & -\frac{1}{F_r^2} \delta_{i3} +\frac{\partial  \tau_{ij}}{\partial x_j} \;\; ,
\end{eqnarray}
\noindent where $R_e=\rho_\ell U_o L_o/\mu_\ell$ is the Reynolds number and $F_r^2 = U_o^2/(g L_o)$ is the Froude number. $g$ is
the acceleration of gravity. $F_i$ is a body force that is used to impose boundary conditions on the surface of the body. $P$ is
the pressure. $\delta_{ij}$ is the Kronecker delta symbol.  As described in \citeasnoun{dommermuth98}, $\tau_{ij}$ is the
subgrid-scale stress tensor. $S_{ij}$ is the deformation tensor:
\begin{eqnarray}
S_{ij} & = & \frac{1}{2} \left( \frac{\partial u_i}{\partial x_j} +\frac{\partial u_j}{\partial x_i} \right) \;\; .
\end{eqnarray}
\noindent $\rho$ and $\mu$ are respectively the dimensionless variable densities and viscosities:
\begin{eqnarray}
\label{density}
\rho(\phi) & = & \lambda + (1 - \lambda) {\rm H} (\phi) \nonumber \\ \mu(\phi) &
= & \eta + (1 - \eta ) {\rm H} (\phi) \;\; ,
\end{eqnarray}
\noindent where $\lambda = \rho_g/\rho_\ell$ and $\eta = \mu_g/\mu_\ell$ are the density and viscosity ratios between air and water.
For a sharp interface, with no mixing of air and water, $H$ is a step function.  In practice, a mollified step function is used to
provide a smooth transition between air and water.

As discussed in \citeasnoun{dommermuth98}, the divergence of the momentum equations (\ref{navi}) in combination with the
conservation of mass (\ref{mass}) provides a Poisson equation for the dynamic pressure:
\begin{eqnarray}
\label{pois} \frac{\partial}{\partial x_i} \frac{1}{\rho} \frac{\partial
P}{\partial x_i} = \Sigma \;\; ,
\end{eqnarray}
\noindent where $\Sigma$ is a source term.  As shown in the next section, the pressure is used to project the velocity onto a
solenoidal field.

\subsection{Numerical Time Integration}

Based on \citeasnoun{sussman03a}, a second-order Runge-Kutta scheme is used to integrate with respect to time the field equations
for the velocity field.  Here, we illustrate how a volume of fluid formulation is used to advance the volume fraction function
(see, for example, \citeasnoun{rider94}).  During the first stage of the Runge-Kutta algorithm, a Poisson equation for the pressure
is solved:
\begin{eqnarray}
\frac{\partial}{\partial x_i} \frac{1}{\rho(\phi^k)} \frac{\partial P^*}{\partial x_i} =\frac{\partial}{\partial x_i} \left(
\frac{u^k_i}{\Delta t}+R_i \right) \;\; ,
\end{eqnarray}
where $R_i$ denotes the nonlinear convective, hydrostatic, viscous, sub-grid-scale, and body-force terms in the momentum equations.
$u^k_i$ and $\rho^k$ are respectively the velocity components at time step $k$.  $\Delta t$ is the time step.  $P^*$ is the
first prediction for the pressure field.

For the next step, this pressure is used to project the velocity onto a solenoidal field. The first prediction for the
velocity field ($u^*_i$) is
\begin{eqnarray}
u^*_i=u^k_i+\Delta t \left( R_i-\frac{1}{\rho(\phi^k)}\frac{\partial P^*}{\partial x_i} \right)
\end{eqnarray}
The volume fraction is advanced using a volume of fluid operator (VOF):
\begin{eqnarray}
\phi^*=\phi^{k}- {\rm VOF}_i \left( u^k_i,\phi^k,\Delta t \right)
\end{eqnarray}
A Poisson equation for the pressure is solved again during the second stage of the Runge-Kutta algorithm:
\begin{eqnarray}
\frac{\partial}{\partial x_i} \frac{1}{\rho(\phi^*)} \frac{\partial P^{k+1}}{\partial x_i}=\frac{\partial}{\partial x_i} \left(
\frac{u^*_i+u^k_i}{\Delta t}+R_i \right)
\end{eqnarray}
$u_i$ is advanced to the next step to complete one cycle of the Runge-Kutta algorithm:
\begin{eqnarray}
u^{k+1}_i=\frac{1}{2} \left( u^*_i + u^k_i +\Delta t \left( R_i -\frac{1}{\rho(\phi^*)}\frac{\partial P^{k+1}}{\partial x_i}
\right) \right) \;\; ,
\end{eqnarray}
and the volume fraction is advanced to complete the algorithm:
\begin{eqnarray}
\phi^{k+1}=\phi^k- {\rm VOF}_i \left( \frac{u^*_i+u^k_i}{2},\phi^{k},\Delta t \right)
\end{eqnarray}

Details of the CLSVOF numerical time-integration procedure are provided in \citeasnoun{sussman03a}.

\subsection{Enforcement of Body Boundary Conditions}

Two different cartesian-grid methods are used to simulate the flow around surface ships. The first technique imposes the no-flux
boundary condition on the body using a finite-volume technique. The second technique imposes the no-flux boundary condition via an
external force field.  Both techniques use a signed distance function $\psi$ to represent the body.  $\psi$ is positive outside the
body and negative inside the body. The magnitude of $\psi$ is the minimal distance between the position of $\psi$ and the surface
of the body. $\psi$ is calculated using a surface panelization of the hull form.   Green's theorem is used to indicate whether a
point is inside or outside the body, and then the shortest distance from the point to the surface of the body is calculated.
Details associated with the calculation of $\psi$ are provided in \citeasnoun{sussman01}.

\subsubsection{\label{sec:body1}Free-slip conditions}

In the finite volume approach, the irregular boundary (i.e. ship hull) is represented in terms of $\psi$ along with the
corresponding area fractions, $A$, and volume fractions, $V$.   $V=1$ for computational elements fully outside the body and $V=0$
for computational elements fully inside the body.  Once the area and volume fractions have been calculated, they are used in the
Poisson equation for the pressure and in the projection of the velocity onto a solenoidal field.  Through the Poisson equation and
the projection operator, the component of velocity that is normal to the ship hull is set to zero.  This corresponds to imposing
free-slip conditions on the hull form.  Details associated with the calculation of the area and volume fractions are provided in
\citeasnoun{sussman01} along with additional references.

\subsubsection{\label{sec:body2}No-slip conditions}

The boundary condition on the body can also be imposed using an external force field.  Based on \citeasnoun{dommermuth98} and
\citeasnoun{sussman01}, the distance function representation of the body ($\psi$) is used to construct a body force in the momentum
equations.  As constructed, the velocities of the points within the body are forced to zero.   For a body that is fixed in a free
stream, this corresponds to imposing no-slip boundary conditions.

\subsection{Interface Capturing}

Two methods are presented in our work for computing ship flows. Both methods use a "front-capturing'' type procedure
for representing the free surface separating the air and water. The first technique is based on the Volume-of-Fluid
(VOF) method, and the second-technique is based on the Coupled volume-of-fluid and level-set method (CLS).

\subsubsection{VOF method}

In our VOF formulation, the free surface is reconstructed from the volume fractions using piece-wise linear polynomials and the
advection algorithm is operator split.  The reconstruction is based on algorithms that are described by \citeasnoun{gueyffier99}.
The surface normals are estimated using weighted central differencing of the volume fractions. A similar algorithm is described by
\citeasnoun{pilliod97}.  Work is currently underway to develop a higher-order estimate of the surface normal using a least-squares
procedure.  The advection portion of the algorithm is operator split, and it is based on similar algorithms reported in
\citeasnoun{puckett97}.

\subsubsection{CLSVOF method}

In the CLSVOF algorithm, the position of the interface is updated through the level-set equation and the volume-of-fluid equation.
After, the level-set function and the volume fractions have been updated, we "couple'' the level-set function to the volume
fractions as a part of the level-set reinitialization step. The level-set reinitialization step replaces the current value of the
level-set function with the exact distance to the VOF reconstructed interface. At the same time, the VOF reconstructed interface
uses the current value of the level-set function to determine the slopes of the piecewise linear reconstructed interface. For more
details of the CLSVOF algorithm, including axisymmetric and three-dimensional implementations, see \citeasnoun{sussman00}.

\subsection{Entrance and Exit Boundary Conditions}

Entrance and exit boundary conditions are required in order to conserve mass and flux.  Two techniques are considered.
The first technique uses a body force, and the second technique uses a special formulation of the pressure.

\subsubsection{Body-force method}

Body forces are used in the momentum equations (see Equation \ref{navi}) and the convection equation for the volume fraction (see
Equation \ref{vof}) to force conservation of flux and mass.   For the velocities, a parallel flow with $(u,v,w)=(-1,0,0)$ is forced
at the entrance and exit.  For the volume fraction, the mean surface elevation is forced to be zero at the entrance and exit. A
similar procedure is used by \citeasnoun{iafrati01} in their level-set calculations of two-dimensional breaking waves over a
hydrofoil. The body-force is prescribed as follows:
\begin{eqnarray}
\label{body}
F_i({\bf x},t) = -F_o T(x) \left( u_i - v_i \right) \;\; ,
\end{eqnarray}
\noindent where $F_o$ is a force coefficient, $v_i=(-1,0,0)$ is the desired velocity field at the entrance and exit,
and $T(x)$ is a cosine taper that smoothly varies from one at the entrance or exit to zero inboard of the entrance or
exit over a distance $L_f$. The formulation for the volume fraction is similar.

\subsubsection{Hydrostatic-pressure method}

At the inflow boundary, the horizontal velocity is set equal to the free-stream velocity and the normal pressure gradient is zero.
At all other side boundaries, the "reduced" pressure is zero and the velocity at the boundary is extrapolated from interior grid cells.
In our computations, we use the "reduced" pressure, $P_r$.  We define $P_r=P-\rho(\phi) g (z-z_o)$, where $z_o$ is the static
free-surface elevation.  The resulting Navier-Stokes equations in terms of $P_r$ are
\begin{eqnarray}
\frac{\partial u_i}{\partial t} + u_j \frac{\partial u_i}{\partial x_j} = -\frac{1}{\rho}\frac{\partial P_r}{\partial x_i}
-\frac{(z-z_o)}{\rho} \frac{\partial \rho g}{\partial x_i}
\end{eqnarray}
\noindent Recall that the density is expressed in terms of a step function (see Equation \ref{density}).  Substitution of
the equation for the density into the preceding equation gives
\begin{eqnarray}
\frac{\partial u_i}{\partial t} + u_j \frac{\partial u_i}{\partial x_j} = -\frac{1}{\rho}\frac{\partial P_r}{\partial x_i}
-\frac{(z-z_o)(1-\lambda) g}{\rho} \frac{\partial H(\phi)}{\partial x_i}
\end{eqnarray}
\noindent The last term is discretized using the same second-order technique used by \citeasnoun{sussman03a} for the
surface-tension term. The last term gives rise to a jump in the reduced pressure of magnitude $(z-z_o)(1-\lambda)g$. By forcing the
reduced pressure to be zero at the walls, over time, the water level at the walls relaxes to $z=z_o$.

\subsection{Initial Transients}

Since VOF simulations are time accurate, there can be problems with starting transients.   As shown by \citeasnoun{wehausen64} and
others, unsteady oscillations can occur in the wave resistance, and by implication the surface elevations, due to starting
transients. There are also starting transients in the buildup of separation and the boundary layer on the hull, but the viscous
time constants are significantly shorter than the wave resistance. The oscillations in the wave resistance occur at a frequency
equivalent to $\omega U_o/g =1/4$ and decay inversely proportional to time. The decay rate is very slow and can lead to solutions
that oscillate for relatively long times. This can problematic if one is trying to reach steady state and also wants to minimize
computer time. For computations presented in this paper, a step function start of the velocity instantaneously jumping to the
free-stream velocity has always been used. Step function starts are easy to initiate in the compute code, but they cause relatively
large transient oscillations.   These very strong initial transients tend to weaken after the body has moved 10 body lengths, but
the weaker oscillations as predicted by \citeasnoun{wehausen64} are still present.  The effects of these transients are reduced by
time averaging.  We note that the oscillations due to the starting transient can be mitigated by reducing the severity of the
startup from a step function to one that is much smoother and slower, which is an option that is currently being investigated.

\section{Results} \label{sec:results}

\subsubsection{NACA 0024 geometry}

The NFA code is used to simulate the flow around a surface-piercing vertical strut moving with constant forward speed.   The water
plane sections of the strut are based on a NACA 0024 section. The numerical results are compared to laboratory measurements that
are reported in \citeasnoun{zhang96}. For the laboratory experiments, the chord length of the model was 1.2m long, and the draft
(1.5m) was sufficiently deep such that at the bottom of the strut the effects of the free surface were minimal.  The Froude number
based on chord length is $F_r=0.55$.

The length, draft, and depth of the computational domain normalized by chord length are respectively 4, 1, and 0.8.   The height of
the computational domain above the mean water line normalized by chord length is 0.2.  The leading edge of the strut is located at
$x=0$ and the trailing edge is located at $x=-1$.  No flux boundary conditions are used on the centerplane of the strut ($y=0$), at
the side of the computational domain ($y=1.0$), the bottom of the computational domain ($z=-0.8$), and the top of the domain
($z=0.2$). Periodic boundary conditions are used along the x-axis at $x=0.3725$ and $x=-3.6275$. The three-dimensional numerical
simulations used $512 \times 128 \times 128=8,388,608$ grid points resulting in a grid spacing along each coordinate axis of
$\Delta x_i=0.0078125$. The time step is $\Delta t=0.00125$, and 3001 time steps have been simulated, which corresponds to 3.75
chord lengths. The number of sub domains along the $x-$, $y-$, and $z-$axes are respectively 32, 8, and 8.  512 CRAY T3E processors
have been used to perform the numerical simulations.  Each time step took approximately 60 seconds per time step.

\begin{figure*}
\begin{center}
\includegraphics[width=4in]{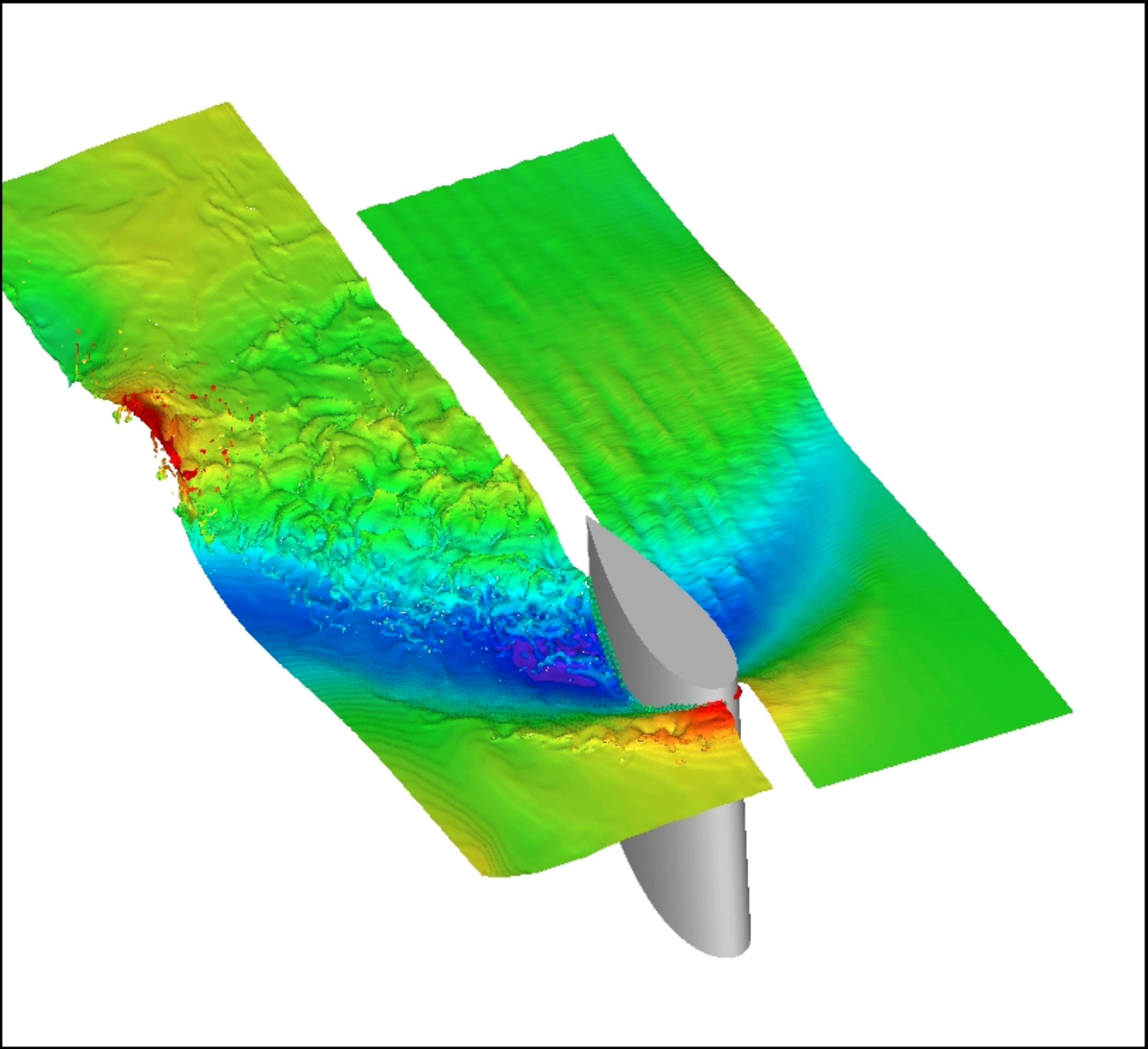}
\end{center}
\caption{\label{comparefoil1} Comparisons to Measurements for Strut Geometry.}
\end{figure*}

Figure \ref{comparefoil1} compares numerical predictions to experimental measurements.  The numerical predictions are shown on the
left side of the strut, and the experimental measurements are shown on the right side of the strut.  The color contours indicate
the free-surface elevation.  Red denotes a wave crest ($\eta=+0.15$) and blue denotes a wave trough ($\eta=-0.15$).  In general,
the agreement between the numerical simulations and the experimental measurements is very good. However, there are some notable
differences. For example, the numerical simulations show more fine-scale detail than the experimental measurements. This is because
the experimental measurements are time-averaged and the numerical simulations show an instantaneous snapshot of the free surface at
$t=3.75$. We also note that unlike the numerical simulations, the measuring device that had been used in the experiments is only
capable of measuring single-valued free-surface elevations.  Another difference between numerical simulations and experimental
measurements occurs away from the strut where the numerical simulations show edge effects due to the smaller domain size that is
used relative to the actual experiments.  Figure \ref{comparefoil1} illustrates that we are able to model the macro-scale features
of the flow associated with the body interacting with the free surface.

\begin{figure*}
\begin{center}
\includegraphics[width=\linewidth]{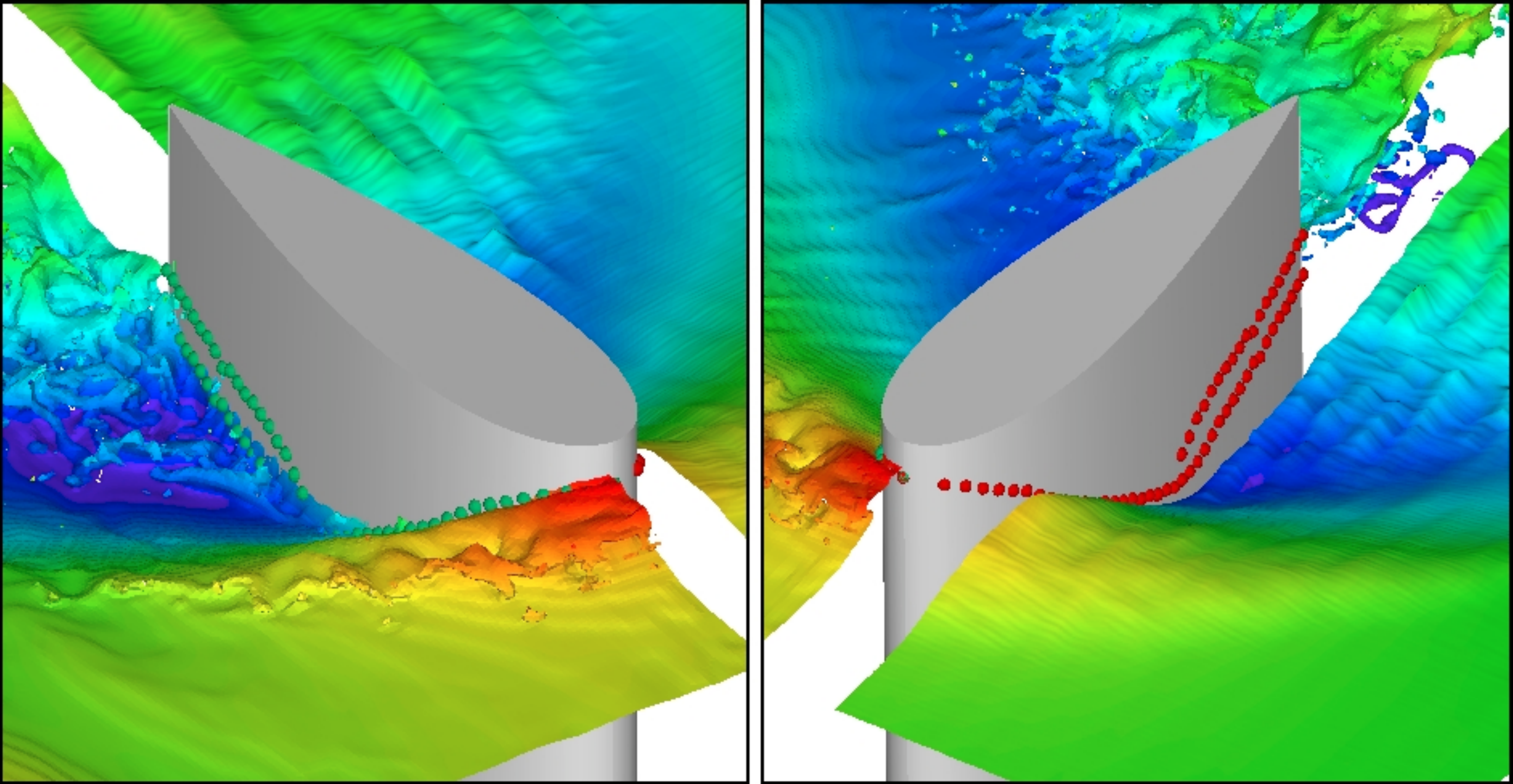}
\end{center}
\caption{\label{comparefoil2} Side Views Looking Down on Strut.}
\end{figure*}

\begin{figure*}
\begin{center}
\includegraphics[width=\linewidth]{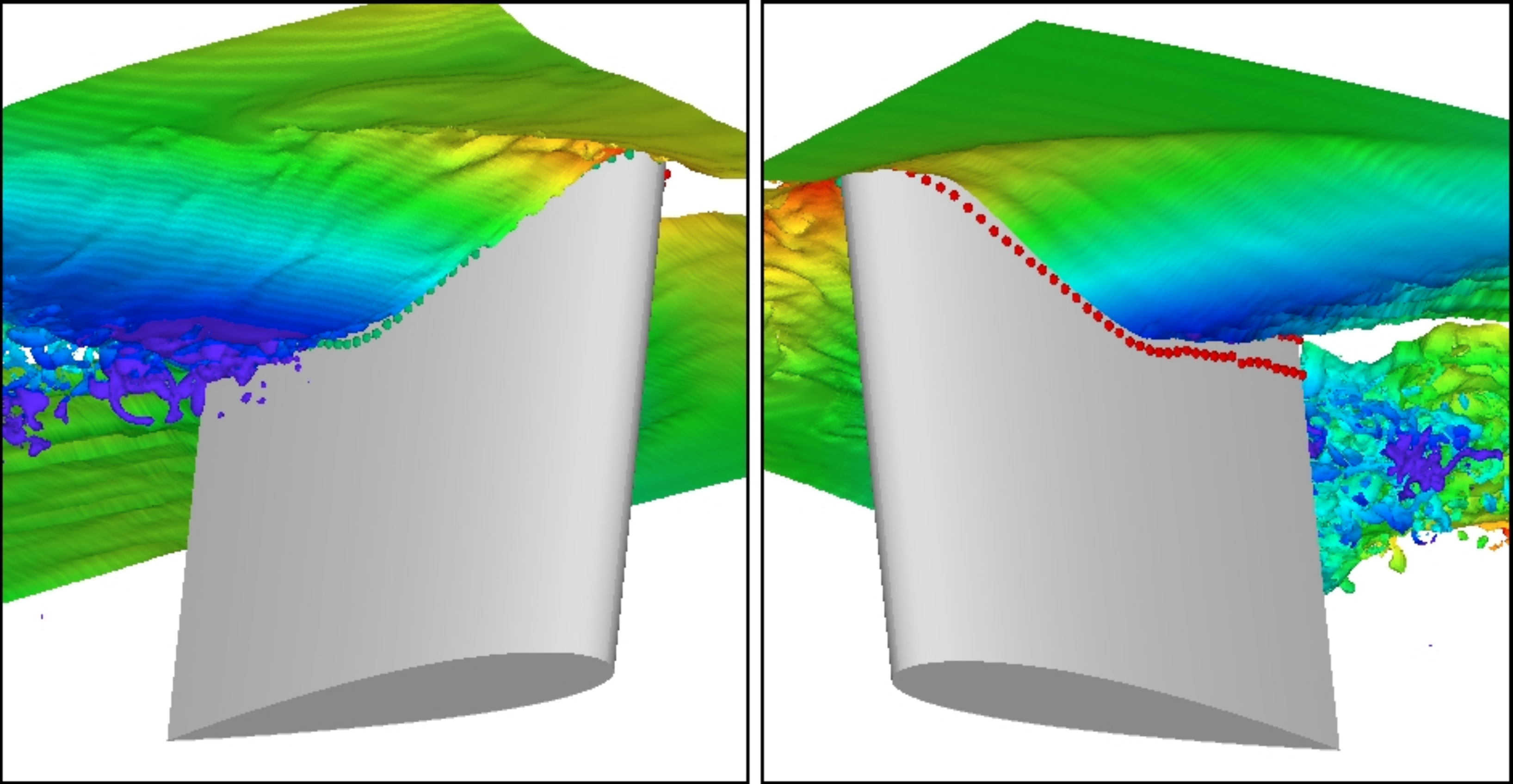}
\end{center}
\caption{\label{comparefoil3} Side Views Looking Up on Strut.}
\end{figure*}

Figures \ref{comparefoil2} and \ref{comparefoil3} show details of the numerical simulations and experimental measurements for two
different views.  The numerical results are shown on the left side of the strut and the experimental measurements are shown on the
right side of the strut.  The red and green dots along the sides of the strut denote experimental measurements of the free-surface
profile.  As before, the color contours indicate the free-surface elevation.  Red denotes a wave crest ($\eta=+0.15$), and blue
denotes a wave trough ($\eta=-0.15$).  Toward the rear of the foil, the dots indicate the upper and lower bounds of the unsteady
rise and fall of the free surface due to flow separation.  Note that in Figure \ref{comparefoil2} and to lesser degree Figure
\ref{comparefoil3}, the experimental contours off of the body do not appear to agree with the experimental profiles on the body.
This is because the contour measurements off of the body could not be performed too close to the body due to limitations associated
with the measuring device. In Figure \ref{comparefoil2}, the numerical simulations show the formation of a spilling breaker and
spray near the leading edge of the strut. Toward the rear of the strut, flow separation is evident.   The numerically predicted
free-surface elevations agree well with the profile measurements in both Figures \ref{comparefoil2} and \ref{comparefoil3}. The
numerical simulations in Figure \ref{comparefoil3} illustrate that air is entrained along the sides of the strut and in the flow
separation zone in the rear.   Additional numerical simulations are in progress to establish the accuracy of the numerical
simulations.

\clearpage

\clearpage
\subsubsection{5415 geometry}

The length, beam, and draft are respectively 5.72m, 0.388m, and 0.248m. The speed is 6.02 knots. Details of the hull
geometry,including the sinkage and trim, are provided at http://www50.dt.navy.mil/5415/.

The length, width, and depth of the computational domain normalized by ship length  are respectively 2, 0.5, and 0.5.  The origin
of our computational domain is taken to be the point at which the unperturbed water intersects the bow of the ship.  The $x$
coordinate at inflow is $x=0.5$ and at outflow, $x=-1.5$. The height of the computational domain above the mean free surface
normalized by ship length is $z=0.5$.  Reduced pressure boundary conditions are used along the sides ($y=\pm0.25$)  and back of the
computational domain ($x=-1.5$).  The free-stream velocity is imposed at the leading edge of the computational domain ($x=0.5$)
with zero pressure gradient. No flux conditions are used at the top and the bottom of the domain ($z=\pm0.5$). The CLSVOF
formulation is used to capture the free-surface interface.   AMR is used locally near the ship hull and the free surface.

Three grid resolutions are considered: low, medium, and high. The low resolution simulation consists of a uniform mesh broken up
into 64 rectangular grid blocks, accounting for $2,097,152$ cells.  The mesh spacing is $\delta=0.0078125$.  The cpu time per time
step is 376 seconds for the low resolution case.  The low resolution simulation was run on 32 processors on an IBM supercomputer
(AIX operating system).

The medium resolution simulation has 64 grid blocks on the coarsest level and 148 grid blocks on the finest level. There are
$5,324,800$ cells for the medium resolution simulation. The cpu time per time step  (32 processors) is 1300 seconds for the medium
resolution case.

The high resolution simulation has 64 grid blocks on the coarsest level, 116 grid blocks on the medium level, and 475 grid blocks
on the finest level accounting for a total of $17,940,480$ cells. The grid spacing on the finest level is $\delta=0.001953125$. The
cpu time per time step (64 processors) is 3000 seconds for the finest resolution case.

\begin{figure}
\begin{center}
\includegraphics[width=3.0in]{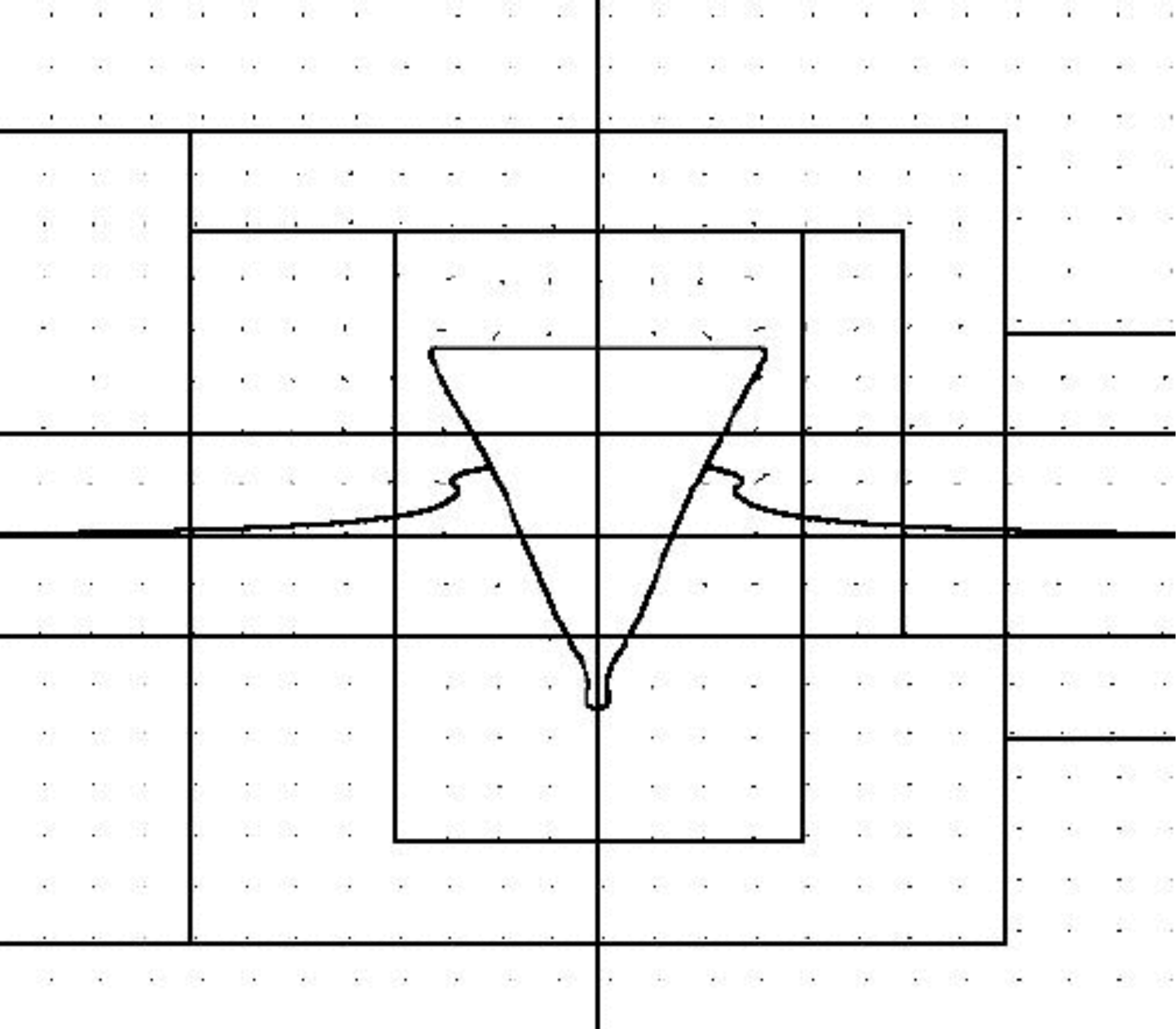}
\end{center}
\caption{\label{slice5415} AMR grid for 5415 at $x=-0.1$.}
\end{figure}

\begin{figure}
\begin{center}
\includegraphics[width=3.0in]{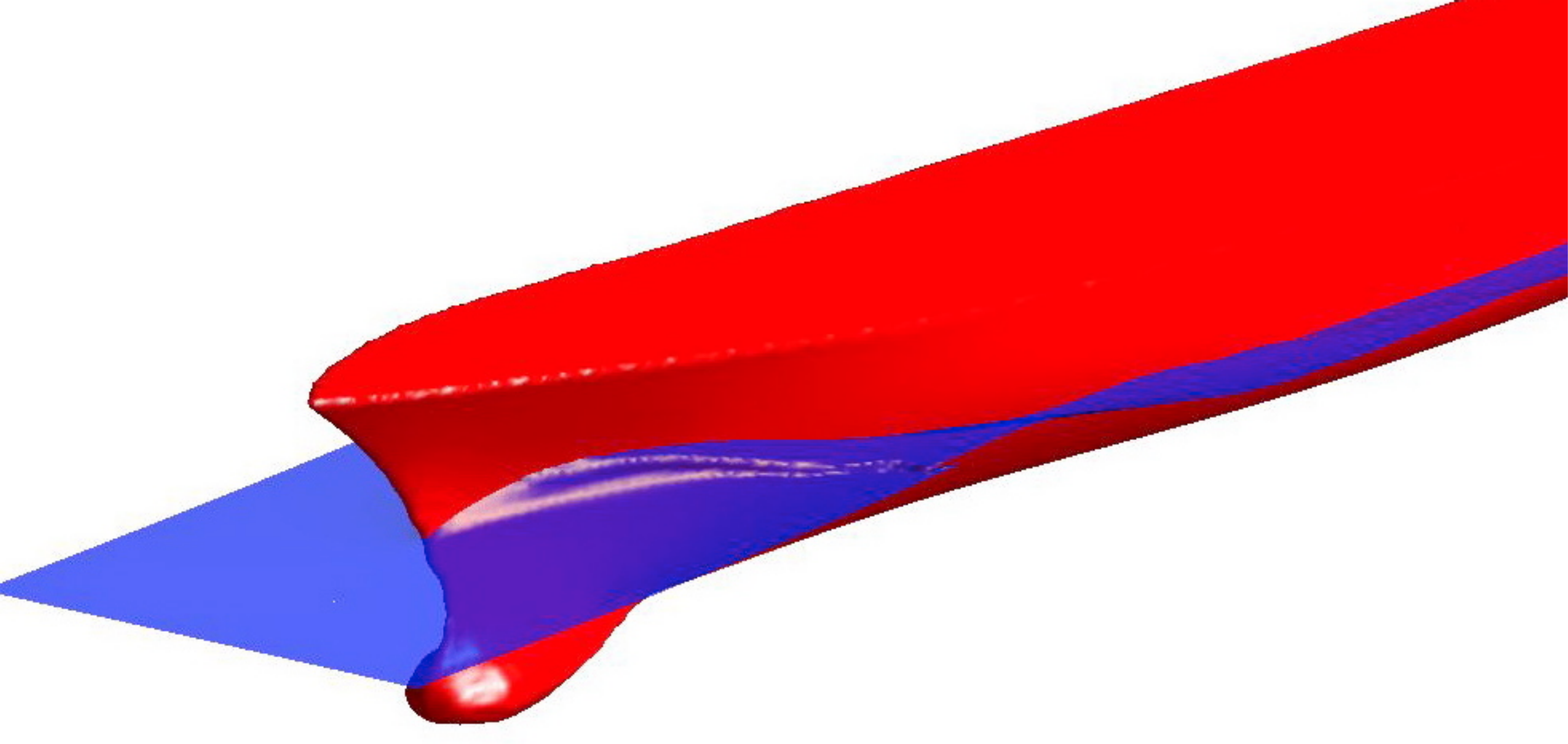}
\end{center}
\caption{\label{bow5415} Bow view of 5415}
\end{figure}

\begin{figure*}
\begin{center}
\includegraphics[width=\linewidth]{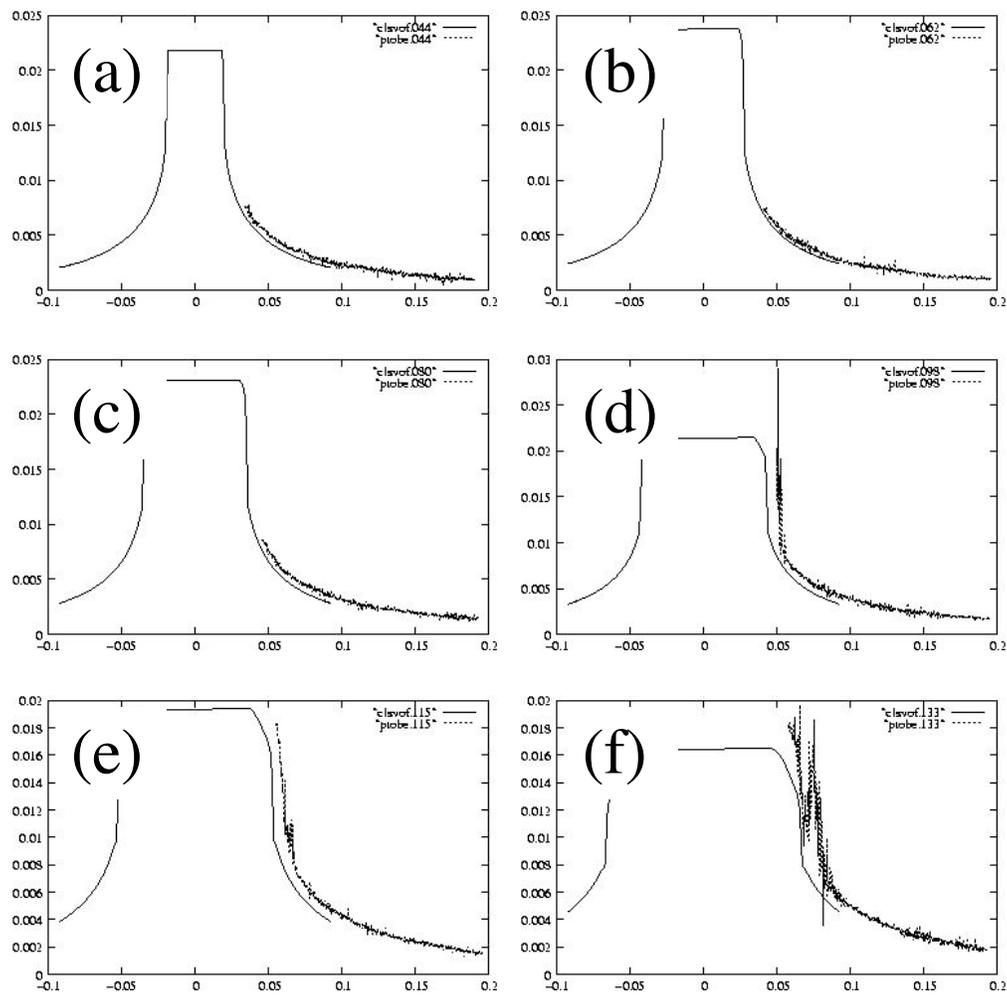}
\end{center}
\caption{\label{whisker5415} CLSVOF predictions compared to whisker-probe measurements for the 5415. (a) $x=-0.044$. (b)
$x=-0.062$. (c) $x=-0.080$. (d) $x=-0.098$. (e) $x=-0.115$. (f) $x=-0.133$.}
\end{figure*}

Figure \ref{slice5415} illustrates the adaptive grid at $x=-0.1$ and $t=1.76$.  Blocks of grid points are clustered near the ship
hull and the free surface.  We note that blocks of grid points are added and deleted over the course of the simulation depending on
resolution requirements.  In analyzing cells advanced per processor, the speed-up due to adding processors or levels of adaptivity
comes to about 70\%.  Figure \ref{bow5415} shows a perspective view of the bow.   The wave overturning that occurs at this Froude
number is clearly visible.  Figure \ref{whisker5415} compares numerical predictions to whisker-probe measurements at various
positions along the $x$-axis.   In general, the agreement between predictions and measurements is very good.  However, the
experiments have fine-scale structure that is not present in the numerics.  Current research is focusing on improving resolution at
the bow by using more levels of AMR.

\clearpage

\clearpage
\subsubsection{Wedge geometry}

The length and draft of the wedge geometry are respectively 35 and 3.5 feet.  The entrance angle is $20$ degrees.  The speed is 6.0
knots.  The Froude number is $F_r=0.3021$. The wedge geometry has a full bow based on the Revelle hull form, and a narrow stern,
based on the bow of the Athena hull form. This enabled the model to be towed in two different directions to investigate the effects
of fullness on the bow wave.  Details of wedge geometry and the towing experiments are provided in \citeasnoun{karion03}.

The length, width, and depth of the computational domain normalized by ship length are respectively 2, 0.8, and 0.5.  The height of
the computational domain above the mean free surface normalized by ship length is 0.1.  The bow is located at $x=0$ and the stern
is located at $x=-1$.  A reflection boundary condition is used on the centerplane ($y=0$) of the wedge.  No-flux conditions are
used on the top ($z=0.1$), bottom ($z=-0.5$), and side ($y=0.8$) of the computational domain.  The flow at the entrance ($x=0.6$)
and exit ($x=-1.4$) are forced to be a parallel flows ($(u,v,w)=(-1,0,0)$) with zero free-surface elevations.

\begin{table}
\begin{center}
{\scriptsize\noindent
\begin{tabular}{|c|c|c|} \hline resolution & grid points & grid spacing  \\ \hline
 coarse & 320 $\times$ 128 $\times$ 96  =  3,932,160 & 0.00625  \\ \hline
 medium & 480 $\times$ 192 $\times$ 144 = 13,271,040 & 0.004166 \\ \hline
 fine & 640 $\times$ 256 $\times$ 192 = 31,457,280 & 0.003125 \\ \hline
\end{tabular}}
\caption{\label{wedge_grid} Grid resolution.}
\end{center}
\end{table}

Three different grid resolutions are used, corresponding to coarse, medium, and fine grid resolutions.  The details with respect to
grid resolution are provided in Table \ref{wedge_grid}. The finest resolution is twice that of the coarsest.  The finest grid
resolution is 0.003125 ship lengths. This would correspond to 31cm for a 100m ship.  In order to resolve large-scale features
associated with spray formation and air entrainment, we believe that grid resolutions less than 10cm are required. Details of the
domain decomposition are provided in Table \ref{wedge_domain}, and the cpu time per time step for each grid resolution are provided
in Table \ref{wedge_cpu}.  Based on these two tables, it can be shown that the cpu time scales linearly with respect to the number
of grid points and the number of processors.  The numerical simulations have been run for 3001 time steps. The time step for each
simulation is $\Delta t=0.002$.

\begin{table}
\begin{center}
\begin{tabular}{|c|c|c|} \hline
resolution & sub-domains & processors  \\ \hline
 coarse     & 20 $\times$  8 $\times$  6 =  960  &  120 \\ \hline
 medium     & 30 $\times$ 12 $\times$  9 =  3240 &  270 \\ \hline
 fine       & 40 $\times$ 16 $\times$ 12 =  7680 &  320 \\ \hline
\end{tabular}
\caption{\label{wedge_domain} Details of the domain decomposition.}
\end{center}
\end{table}

\begin{table}
\begin{center}
\begin{tabular}{|c|c|c|} \hline
resolution & cpu time per time step (sec) \\ \hline
 coarse     & 30.9 \\ \hline
 medium     & 47.5 \\ \hline
 fine       & 94.8 \\
\hline
\end{tabular}
\caption{\label{wedge_cpu} CPU speed.}
\end{center}
\end{table}

\begin{figure}
\begin{center}
\includegraphics[height=3.5in]{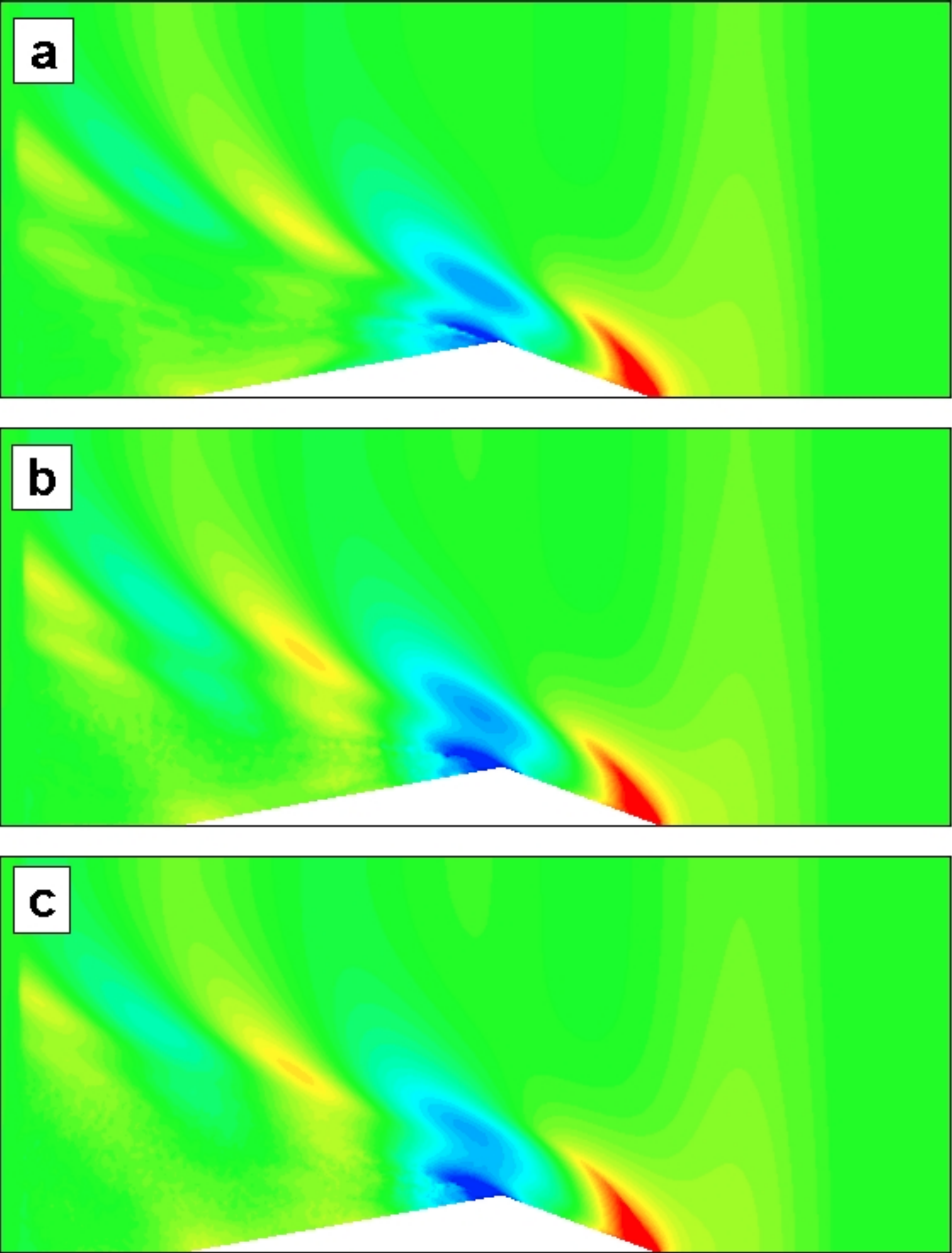}
\end{center}
\caption{\label{mean_vof} Mean surface elevation. (a) Coarse. (b) Medium (c) Fine.}
\end{figure}

The mean surface elevations for each grid resolution are shown in Figure \ref{mean_vof}. The flow is from right to left.  The color
contours indicate the free-surface elevation.  Red denotes a wave crest ($\eta=+0.025$), and blue denotes a wave trough
($\eta=-0.025$). The mean position of the free surface is calculated from the volume fraction averaged over time from $t=4$ to
$t=6$. Based on this time average, the mean position of the free surface is defined as the 0.5 isosurface.  As grid resolution
increases, the bow wave becomes steeper.  In addition, the trough in the flow separation region at the corner of the wedge gets
deeper.   The wave rays also become more distinct.

\begin{figure}
\begin{center}
\includegraphics[height=3.5in]{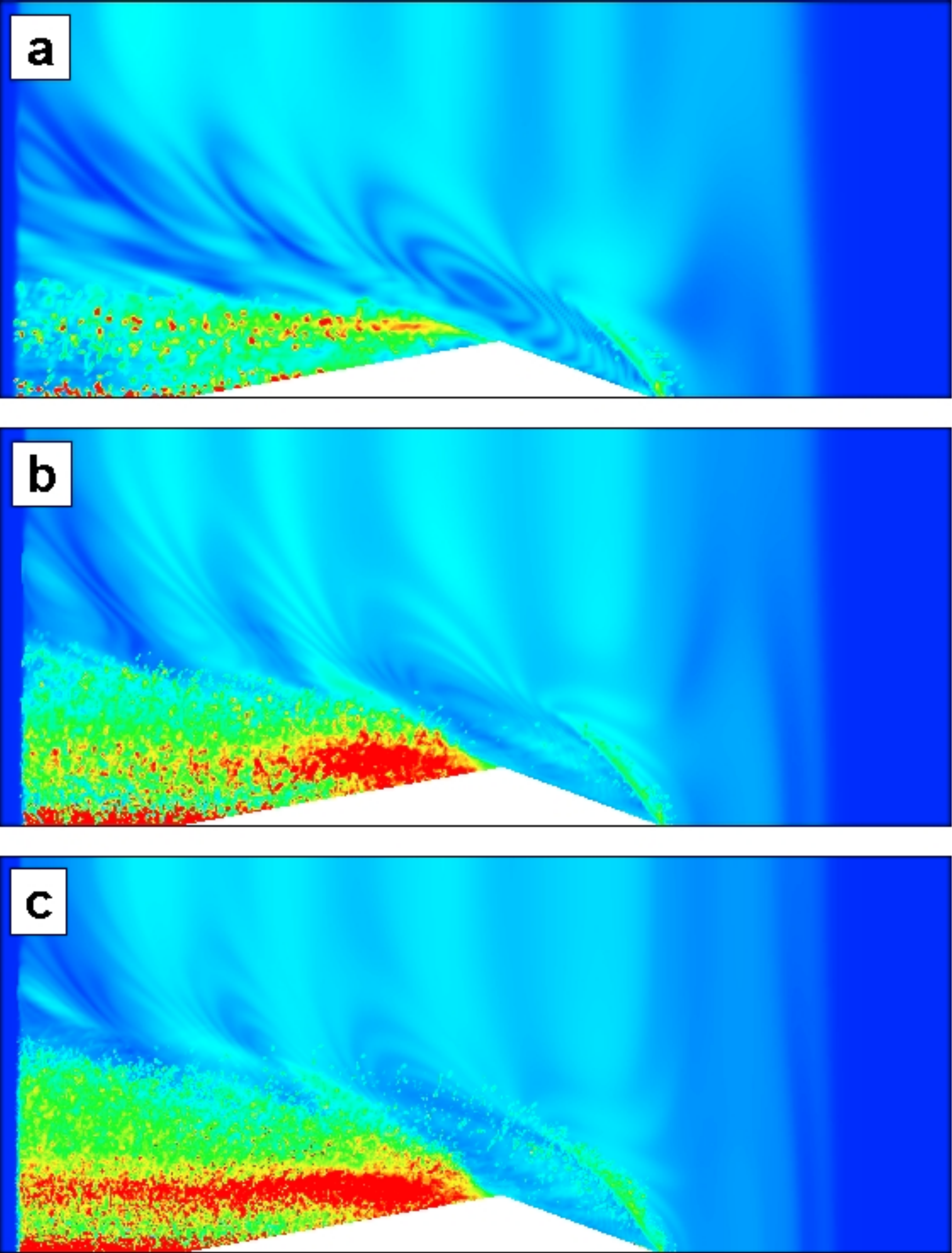}
\end{center}
\caption{\label{rms_vof} RMS surface fluctuations. (a) Coarse. (b) Medium (c) Fine.}
\end{figure}

The correlation coefficient between the coarse and medium resolution simulations is 0.94, and the correlation coefficient between
the medium and fine resolution simulations is 0.99.  The rms differences between the coarse and medium resolution simulations is
$1.83\times10^{-3}$, and the rms differences between the medium and fine resolution simulations is $6.68\times10^{-4}$.   This
demonstrates that the prediction of mean quantities is converging. Increasing grid resolution also improves resolution of
small-scale fluctuations as shown in Figure \ref{rms_vof}.

\begin{figure}
\begin{center}
\includegraphics[height=3.5in]{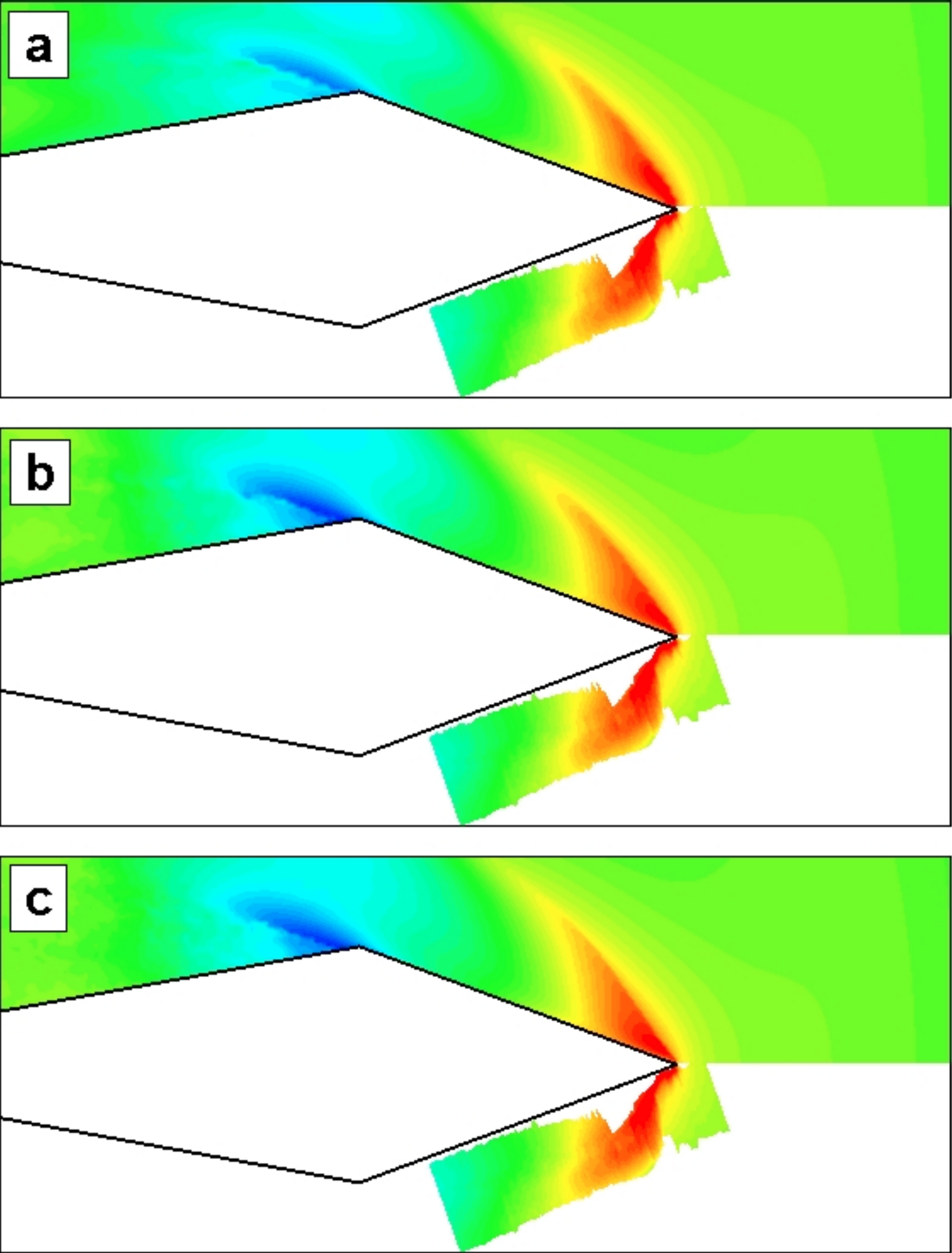}
\end{center}
\caption{\label{comparewedge} Comparisons to QViz Measurements for Wedge Geometry. (a) Coarse. (b) Medium (c) Fine.}
\end{figure}

The rms surface fluctuations for each grid resolution are shown in Figure \ref{rms_vof}.  The color contours indicate the magnitude
of the free-surface fluctuations. Red denotes the maximum rms fluctuations ($\tilde{\eta}=0.011$), and blue indicates regions where
there are no fluctuations.  The rms fluctuations are calculated by taking the square root of the variance of the vertical offset
where the phase changes from air to water.  The regions where phase changes occur include droplets of fluid above the mean position
of the free surface and bubbles of air beneath the mean position of the free surface.   The statistics are calculated from $t=4$ to
$t=6$.  A histogram analysis indicates that phase changes are dominated by small-scale fluctuations on the mean position of the
free surface. This corresponds to roughening of the free surface.   The greatest fluctuations in the free-surface elevation occur
along the centerline of the wedge, in the flow separation region behind the corner of the wedge, and along the front face of the
bow wave. Comparisons of coarse, medium, and fine resolutions show that fluctuations increase as the grid resolution increases.
Interestingly, the finest resolution simulation shows that the rms fluctuations increase in extent slightly off of the center plane
on the front face of the bow wave. Based on photographs of the experiments \cite{karion03}, this is a region where the bow wave
overturns.  This effect is also evident in the measurements shown in Figure \ref{comparewedge}.

The mean free-surface elevation for each grid resolution compared to laboratory measurements are shown in Figure
\ref{comparewedge}.  Numerical predictions are plotted in the top portion of each graph.   Quantitative Visualization (QViz)
measurements are plotted in the bottom portion of each graph.  QViz uses a laser sheet to illuminate the free surface. A video
camera is used to capture snapshots, which are then digitally processed. Additional details of the QViz measurements are provided
in \citeasnoun{karion03}. The color contours indicate the free-surface elevation. Red denotes a wave crest ($\eta=+0.035$), and
blue denotes a wave trough ($\eta=-0.035$).  As before, the mean position of the free surface is calculated from the volume
fraction averaged over time from $t=4$ to $t=6$.  For these figures, $-0.687 \leq x \leq 0.284$ and $-0.21 \leq y \leq 0.21$.  The
correlation coefficients between the measurements and the predictions for the coarse, medium, and fine simulations are respectively
0.951, 0.954, and 0.957.   Since the QViz instrument measures from the top down, we also consider the correlation between the
experimental data and the predictions of the mean plus the rms fluctuations.  In this case, the correlations improve to 0.950,
0.958, and 0.960 for respectively the coarse, medium, and fine simulations.

\begin{figure}
\begin{center}
\includegraphics[height=3.5in]{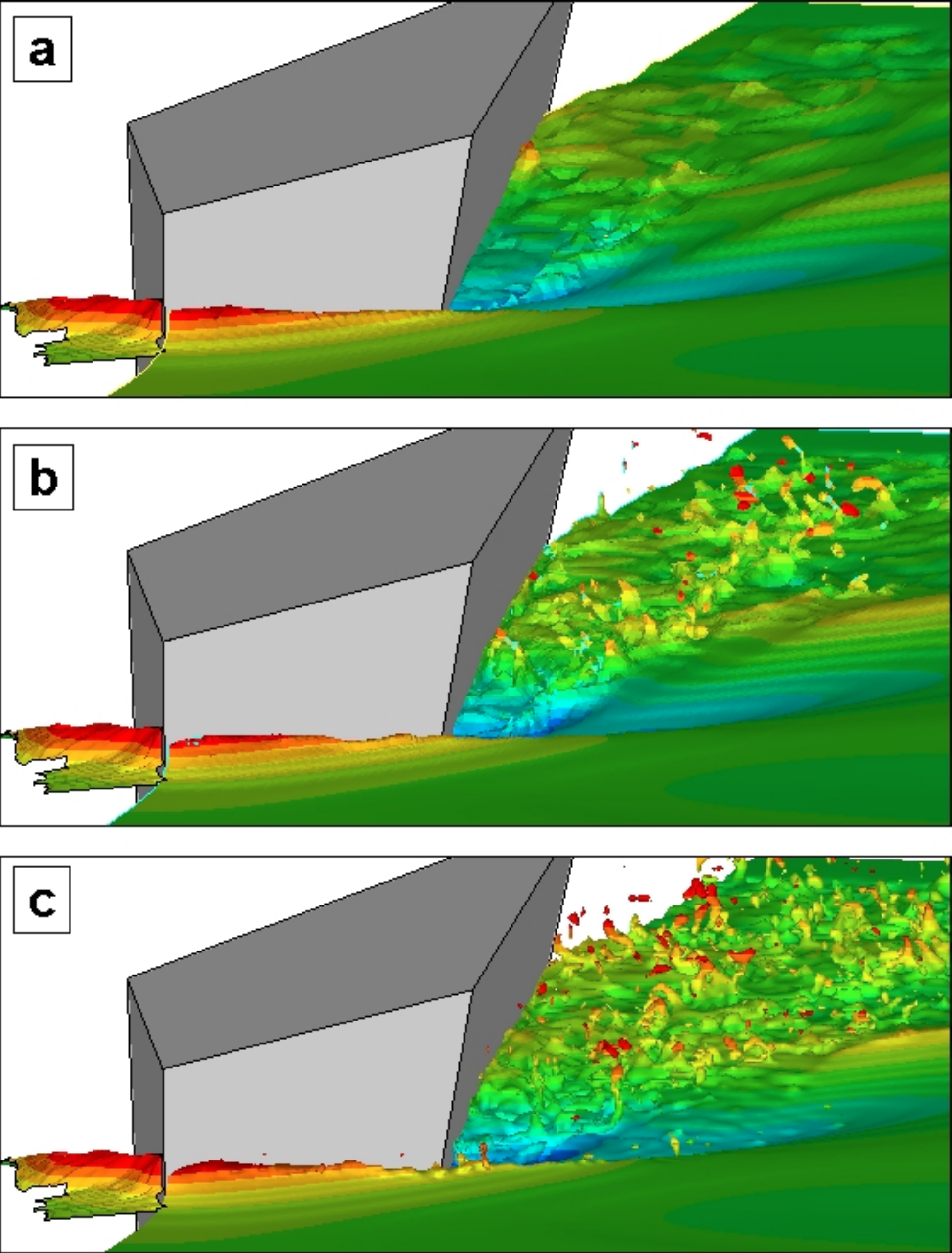}
\end{center}
\caption{\label{perspective} Perspective views of wedge. (a) Coarse. (b) Medium (c) Fine.}
\end{figure}

Perspective views of the free-surface deformation for each grid resolution are shown in Figure \ref{perspective}.  The color
contours indicate the free-surface elevation.  Red denotes a wave crest ($\eta=+0.03$), and blue denotes a wave trough
($\eta=-0.03$).  QViz measurements are plotted on the left side of each figure.   Snap shots of the free surface at time $t=6$ for
each grid resolution are plotted on the right side of each figure.  As grid resolution increases, the fragmentation of the free
surface also increases.   We conjecture that the large-scale break up of the free surface is dominated by inertial effects and that
the effects of surface tension are only important at the very smallest scales.  Work is currently in progress to test this
assertion.

\section{Conclusions}

With sufficient resolution, interface capturing methods are capable of modelling the formation of spray and the entrainment of air.
Based on comparisons to other VOF formulations that are not reported here, second-order-accurate formulations such as those used in
the NFA and CLSVOF codes are desirable because first-order schemes tend to inhibit wave breaking.  A major benefit of our
cartesian-grid formulations relative to body-fitted formulations is that second-order VOF formulations are easier to develop.

In terms of future research, an AMR capability is currently being developed for the NFA code.  For our second-order VOF
formulation, a key issue is mass conservation and surface reconstruction along boundaries where grid resolution changes. Various
methods are also being investigated to reduce initial transients.  One method slowly ramps up the free-stream velocity, which is
similar to how a towing-tank carriage operates.  We are also continuing development of techniques for processing VOF datasets to
improve understanding and modelling of wave breaking.

\section{Acknowledgements}

This research is supported by ONR under contract numbers N00014-04-C-0097 and N00014-02-C-0432. Dr. Patrick Purtell is the program
manager. The second author is supported in part by the NSF Division of Mathematical Sciences under award number DMS 0108672 with
Thomas Fogwell as program manager and by ONR under contract number N00014-02-C-0543 with Judah Goldwasser as program manager. The
numerical simulations have been performed on the Cray T3E at the U.S. Army Engineering Research and Development Center.

\bibliography{25onr}
\bibliographystyle{25onr}

\end{document}